\documentclass[journal]{IEEEtran}

\usepackage{xcolor,soul,framed} 

\colorlet{shadecolor}{yellow}
\usepackage[pdftex]{graphicx}
\graphicspath{{../pdf/}{../jpeg/}}
\DeclareGraphicsExtensions{.pdf,.jpeg,.png}

\usepackage[cmex10]{amsmath}
\usepackage{amssymb}

\usepackage{algorithm,algcompatible,amsmath}
\usepackage{array}
\usepackage{mdwmath}
\usepackage{mdwtab}
\usepackage{eqparbox}
\usepackage{url}

\usepackage{subcaption}
\usepackage[]{graphicx}  
\usepackage{nccmath}
\usepackage{cite}
\usepackage{multirow}
\usepackage{caption}
\usepackage{booktabs}
\usepackage{verbatim}
\usepackage{bm}
\usepackage{float}
\newcolumntype{C}{>{\centering\arraybackslash}X} 

\begin{document}
\bstctlcite{IEEEexample:BSTcontrol}
    \title{Safety-Enhanced Self-Learning for Optimal Power Converter Control}
  \author{Yihao Wan,~\IEEEmembership{Student Member,~IEEE,} Qianwen Xu,~\IEEEmembership{Member,~IEEE,} and
       Tomislav Dragičević,~\IEEEmembership{Senior Member,~IEEE}\vspace{-9mm}


  \thanks{Yihao Wan and Tomislav Dragičević are with the Department of Wind and Energy Systems, Technical University of Denmark, Copenhagen, Denmark (e-mails: wanyh@dtu.dk, tomdr@dtu.dk). }
  \thanks{Qianwen Xu is with the School of Electrical Engineering and Computer Science, KTH Royal Institute of Technology, Stockholm, Sweden (email: qianwenx@kth.se).}
  
  }  

\markboth{
}{Roberg \MakeLowercase{\textit{et al.}}: High-Efficiency Diode and Transistor Rectifiers}

\maketitle

\begin{abstract}
Data-driven learning-based control methods such as reinforcement learning (RL) have become increasingly popular with recent proliferation of the machine learning paradigm. These methods address the parameter sensitiveness and unmodeled dynamics in model-based controllers, such as finite control-set model predictive control. RL agents are typically utilized in simulation environments, where they are allowed to explore multiple "unsafe" actions during the learning process. However, this type of learning is not applicable to online self-learning of controllers in physical power converters, because unsafe actions would damage them. To address this, this letter proposes a safe online RL-based control framework to autonomously find the optimal switching strategy for the power converters, while ensuring system safety during the entire self-learning process. The proposed safe online RL-based control is validated in a practical testbed on a two-level voltage source converter system, and the results confirm the effectiveness of the proposed method.
\end{abstract}

\vspace{-4mm}
\vspace{-4mm}

%
\IEEEpeerreviewmaketitle


\section{Introduction}
\IEEEPARstart{R}{einforcement} 
learning (RL) has gained increasing attention for applications in the power electronics field due to its model-free and self-learning characteristics. Compared with conventional cascaded linear control strategies that suffer from slow dynamics and finite control-set model predictive control (FCS-MPC) that has fast response and straightforward design but rely heavily on accurate parameters and the established model \cite{dragivcevic2020advanced}, the learning-based controllers can simultaneously mitigate the dependence on the precise system model and achieve fast and accurate control performance. 

Different RL-based controllers have been proposed for converters. In \cite{gheisarnejad2020novel}, an RL-based controller is incorporated into the sliding mode observer to stabilize the buck-boost converter and improve the voltage regulation under constant power load. In addition, different RL algorithms are used to optimize the shift angle of the triple phase shift modulation in dual active bridge converters to minimize reactive power\cite{tang2021artificial}, power losses\cite{tang2022deep}, and current stress \cite{zeng2022autonomous}. 

However, a pivotal challenge in RL lies in striking the balance between exploration and exploitation, often leading to protracted convergence time for RL agents in finding the optimal policy. Moreover, the physical limitations are not ensured during training sessions, especially in the initial stages involving random exploration, thereby limiting its practical application within the domain of online self-learning in power electronics. Recent advancements have embraced the safe RL paradigm to address these issues \cite{weber2023safe, xia2022safe, chen2022physics}. Existing safety frameworks for RL predominantly target optimization and power systems, accomplished either through direct action evaluation against constraints \cite{weber2023safe}, training a safety model correlating actions with constraint functions \cite{xia2022safe}, or substituting unsafe actions with those derived from physics-related functions \cite{chen2022physics}, etc. Nevertheless, there appear to be research gaps concerning safe learning-based control for power converters.

Motivated by these, this letter proposes a novel safety-enhanced self-learning approach for optimal power converter control. The proposed safe learning framework combines model-based and data-driven learning-based approaches by incorporating a computationally efficient MPC-based safety policy into the learning-based controller. The proposed safe learning-based controller can achieve comparable control performance to the conventional FCS-MPC, guarantee system safety throughout the learning process, and significantly improve learning efficiency. 
\vspace{-2mm}
\section{Unsupervised online safe reinforcement learning based controller}
\begin{figure}[H]
	\begin{center}
		\includegraphics[width=0.65\linewidth]{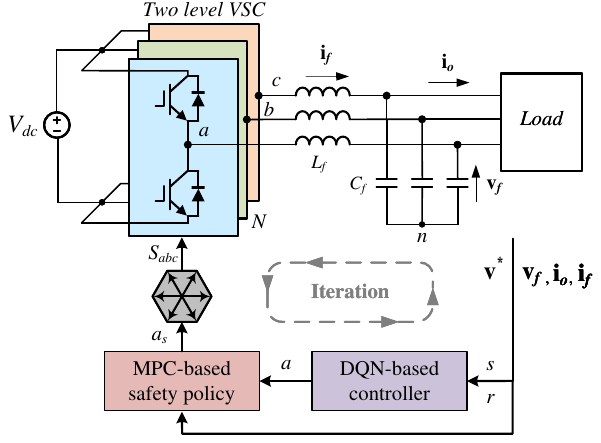}\\
		\caption{Schematic of the safe learning-based control for VSC.}
		\label{schematic}
	\end{center}
        \vspace{-8mm}
\end{figure}

The studied system is shown in Fig. \ref{schematic}, where a two-level voltage source converter (VSC) interfaces the load and DC source. In particular, eight possible switching combinations align with different voltage vectors. The conventional FCS-MPC predicts the converter behaviors for those switching combinations based on a model discretized with a sampling time $T_s$, shown below, where a cost function is used to select the optimal switching states\cite{dragivcevic2017model}. 
\begin{equation}
\small
 \begin{bmatrix} \mathbf i_f (k+1)\\\mathbf v_f (k+1)\end{bmatrix} = \begin{bmatrix} -\frac{R_f T_s}{L_f}& -\frac{T_s}{L_f}\\ \frac{T_s}{C_f}& 0\end{bmatrix} \begin{bmatrix} \mathbf i_f (k)\\\mathbf v_f (k)\end{bmatrix} + \begin{bmatrix} \frac{T_s}{L_f}& 0\\ 0& -\frac{T_s}{C_f}\end{bmatrix} \begin{bmatrix} \mathbf v_i (k)\\\mathbf i_o (k)\end{bmatrix}
\label{discrete}
\vspace{-5mm}
\end{equation}
\subsection{Problem formulation for the RL}\label{formulation}
RL paradigm for learning optimal converter control is formulated as a Markov Decision Process (MDP). Due to the limited number of switching combinations, the DQN algorithm with a discrete action space is employed. The proposed RL-based controller aims to find the optimal switching state selection policy and regulate the VSC by merely interacting with the converter system. As shown in Fig. \ref{schematic}, at each discrete time step $t$, the DQN agent receives the measurements $s_t$, and takes an action $a_t$ according to the policy $\pi$. At time step $t$+1, the system shifts to a new state $s_{t+1}$, and the agent receives a reward $r_t$ for the transition. The agent takes actions according to the maximum $Q$ value with $\epsilon$-greedy strategy, described as
\begin{equation}
\small
a_t = \begin{cases}
 \rm arg\,\mathop{max}\it\limits_a\it Q_{\pi}(s_t,a) ,\quad & \rm with\;probability\;\epsilon; \\
\rm a\;random\;action,\quad & \rm otherwise.
\end{cases} 
\label{action_selection}
\end{equation}

In DQN, a deep neural network is employed to evaluate the $Q$ value for each state-action pair. During the training session, transition sequences $(s_t,a_t,r_t,s_{t+1})$ are stored in a replay memory, where a minibatch of the tuples is randomly selected to train the neural network via stochastic gradient descent. In addition, to enhance the stability and convergence of training the network with parameter $\theta$, a target network $Q_{\pi}^{'}$ synchronizing its separate network parameter $\theta^{'}$ with the $\theta$ for fixed time step is introduced. The $Q$ value is updated as
\begin{equation}
		\small
        y_j = \left\{
			\begin{array}{@{}ll@{}} 
			r_j,\rm if\;episode\;terminates\;at\;step\;\textit j+\rm1; \\ 
			r_j + \gamma \rm max \it Q^{'}_{\pi}(s_{j+1},a^{'}|\theta^{'}), \rm otherwise. 
			\end{array}\right.
	   \label{qvalue} 
\end{equation}

The loss function for updating the $Q$-network parameter is 
\begin{equation}
		\small
		L(\theta) = \mathbb{E}\left[(y_t - Q_{\pi}(s_t,a_t|\theta))^2\right]
	\label{critic_loss}
\end{equation}
\vspace{-10mm}
\subsection{RL-based controller design}\label{formulation}
Using a reward formulated to incentivize the RL agent, the proposed RL-based controller learns autonomously to select the optimal switching state based on the input measurements, emulating the switching state selection strategy of FCS-MPC.

\subsubsection{State and action sets}
To achieve similar control performance to the FCS-MPC, the input states include reference voltage ($v^{*}_{\alpha}$, $v^{*}_{\beta}$), capacitor voltage deviations ($\triangle v_{f\alpha}$, $\triangle v_{f\beta}$), filter currents ($i_{f\alpha}$, $i_{f\beta}$), and the previous action, i.e., the voltage vector $x(k-1)$ from the previous sampling period. The action space consists of the number of voltage vectors as $[1:1:7]$, matching the switching combinations, respectively.

\subsubsection{Reward function design}
The reward function should be finite for feasible and efficient training. To emulate the optimal switching states selection strategy in FCS-MPC and regulate the AC voltage, the reward function is thus formulated as

\begin{equation}
\small
r = -[(v^*_{\alpha}-v_{f\alpha})^2 + (v^*_{\beta}-v_{f\beta})^2] 
\label{reward_imitation}
\end{equation}
\vspace{-10mm}
\subsection{Safety policy for the online RL}\label{safety policy}
Practically, the converter-side current should always be limited within an allowable range to ensure hardware safety. Therefore, the actions taken by the DQN agent should be considered for the sake of the system's safety. To achieve this, a safety framework based on a computational light single-step prediction illustrated in Fig. \ref{schematic} is proposed. 

Specifically, at each training step, the action taken by the RL agent from (\ref{action_selection}) is input to the MPC-based safety block, which performs one-step prediction for the current  $i_f (k+1)$ based on equations in (\ref{discrete}). It will also evaluate whether the corresponding switching states would lead to overcurrent as
\begin{equation}
\small
||i^a_{f\alpha\beta}(k+1)||_2 \leq i_{max}
\label{current_limit}
\end{equation}
\vspace{-4.5mm}

The unsafe actions causing overcurrent will be abandoned. Instead, safe switching states from the MPC-based safety block will be used to ensure the safe operation of the system at all times. Notably, the safety framework is not to find optimal actions but to guide the RL agent to take safe actions. Meanwhile, the agent will gradually converge within the safe learning space, bypassing the safety block. In this way, the safety framework also narrows the learning spaces by avoiding unsafe actions, improving the learning efficiency. In short, the proposed safety-enhanced self-learning optimal controller combines a computationally light single-step prediction to exclude unsafe actions and narrow the learning space with an RL-based learning framework. 
\vspace{-2mm}
\section{Experimental Results and Discussion}
\begin{figure}[H]
	\begin{center}
		\includegraphics[width=0.66\linewidth]{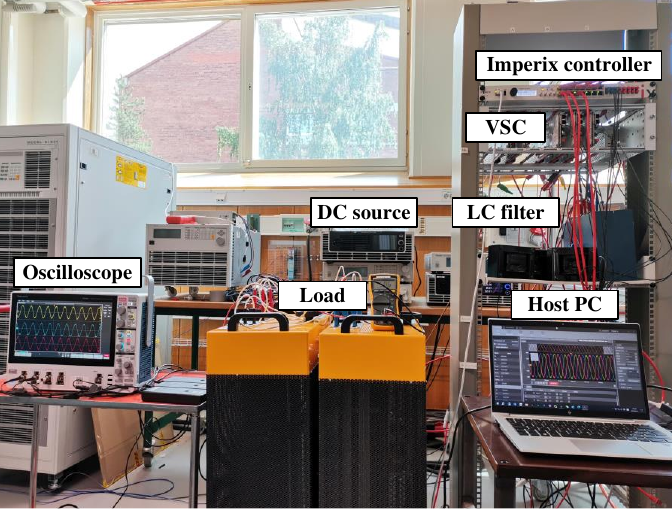}\\
		\caption{Experimental setup of a two-level VSC system.}
		\label{exp_setup}
	\end{center}
    \vspace{-6mm}
\end{figure}

This section presents experimental validations for the proposed safe online RL-based control framework. The experimental setup in Fig. \ref{exp_setup} aligns with the configuration in Fig. \ref{schematic}. The DC link voltage is 520 V, the $LC$ filter is 2.5 $mH$ and 30 $\mu\,F$, the load is 50 $\Omega$, the reference voltage is 200 V with frequency 50 Hz, the sampling time $T_s = 20\,\mu s$, and $i_{max}$ is set as 20 A. In particular, the DQN agent is trained via an edge device, where a simulation model matching the testbed parameters is built in MATLAB/SIMULINK. The training in simulation on the edge device can stay the same as the real converter system, where the agent could also be trained online safely with the proposed method. Afterward, the online safe RL-based controller for VSC is transferred from edge devices to the practical setup using the IMPERIX control platform. 

\begin{figure*}
	\centering
	\begin{subfigure}{0.38\textwidth}
		\centering
		\includegraphics[width=\linewidth]{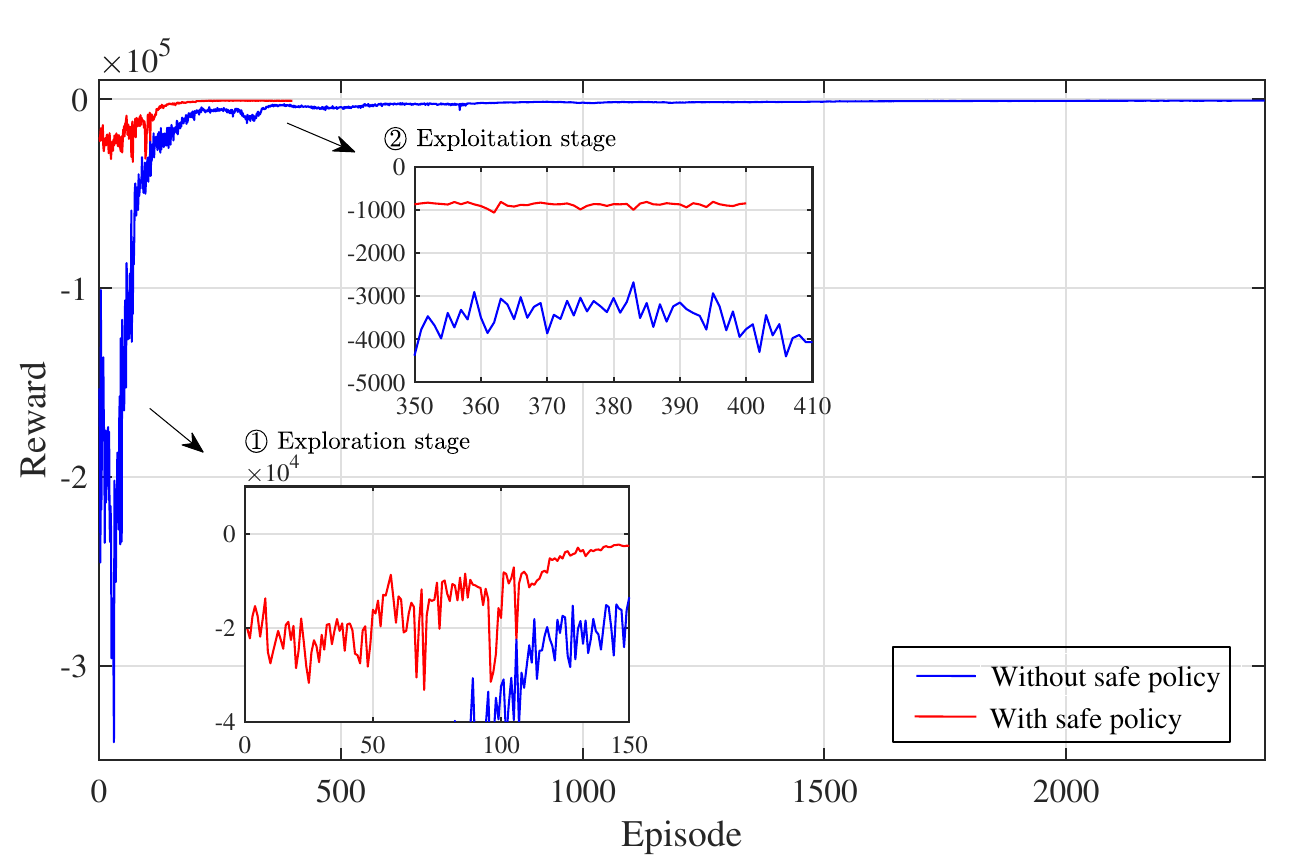}  
		\caption{Average accumulated reward}
		\label{reward}
	\end{subfigure}	
	\begin{subfigure}{0.24\textwidth}
		\centering
		\includegraphics[width=\linewidth]{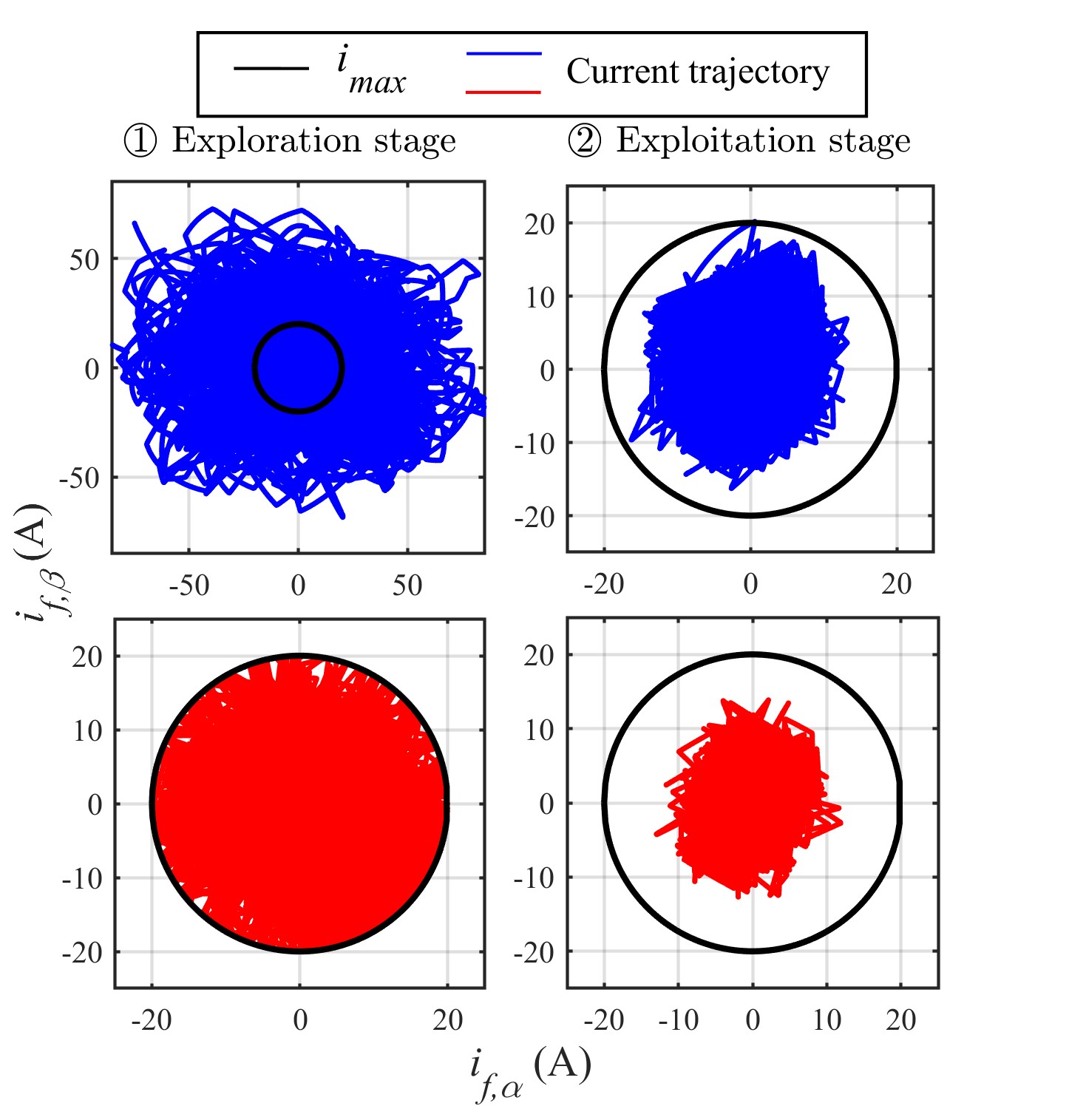}  
		\caption{Converter-side current trajectory}
		\label{current_tj}
	\end{subfigure}
	\caption{Training process of the proposed safety-enhanced self-learning optimal controller.}
	\label{train_results}
    \vspace{-6mm}
\end{figure*}

The training process of the DQN-based controller is shown in Fig. \ref{train_results}. Starting from exploration, it can be observed from Fig. \ref{reward} that the agent without the safe policy takes actions randomly, resulting in significant deviations of the output voltage from the reference voltage, causing a sharp drop in the average accumulated reward. In addition, the converter-side current trajectory shown in Fig. \ref{current_tj} exceeds the maximum current, which would damage the converter. As the training episodes progress, the RL agent learns and converges within the safe learning region. Conversely, the proposed safe RL agent explores the safe learning space to constrain the current within the physical limit throughout the training process. The accumulated reward is much higher, and the agent quickly converges to the optimal policy. In this way, the accumulated average reward for the two training methods converges to the same level, and the RL agent learns the optimal switching strategy for the VSC, emulating the conventional FCS-MPC. 

\begin{figure}
	\centering
	\begin{subfigure}{0.28\textwidth}
		\centering
		\includegraphics[width=\linewidth]{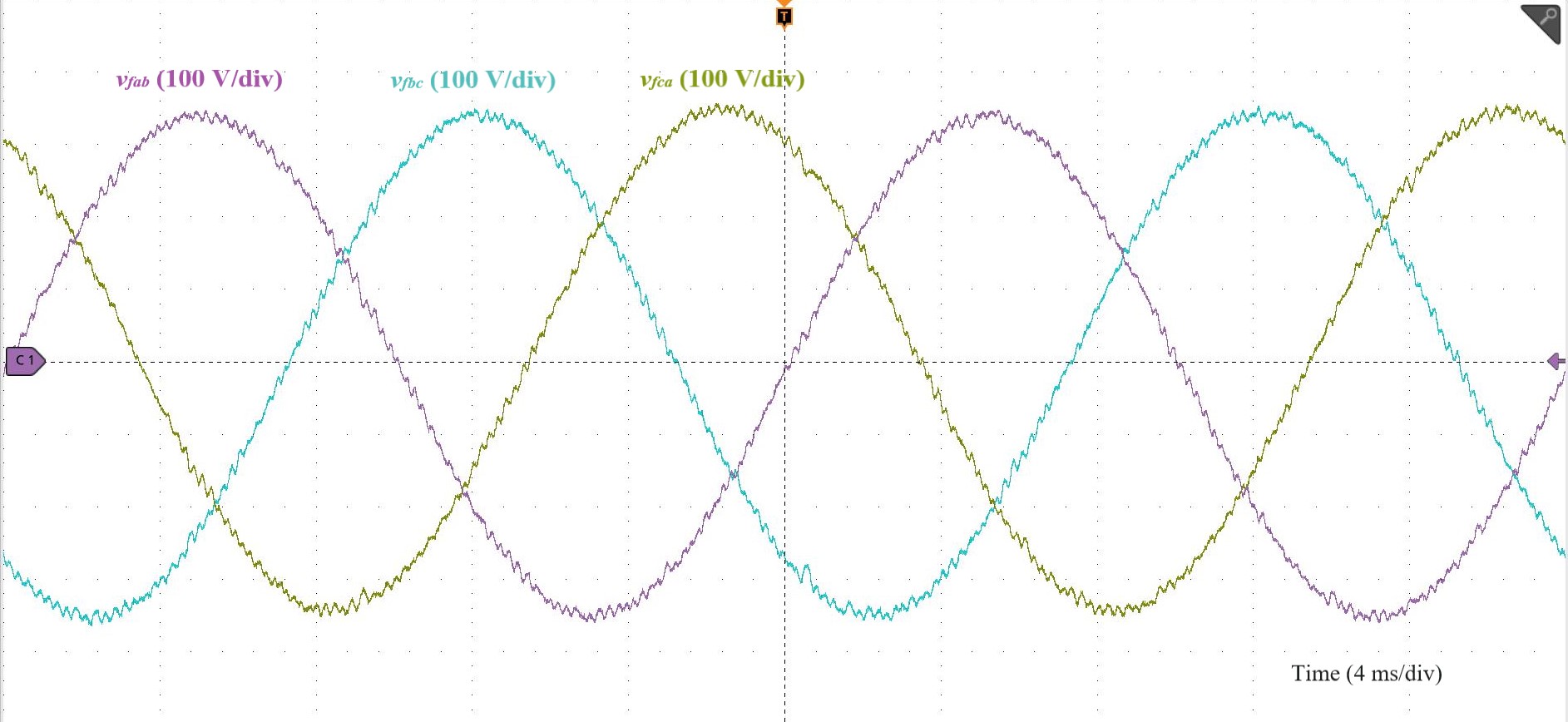}  
		\caption{Conventional FCS-MPC. THD = 1.97\%.}
		\label{mpc}
	\end{subfigure}
	
	\begin{subfigure}{0.28\textwidth}
		\centering
		\includegraphics[width=\linewidth]{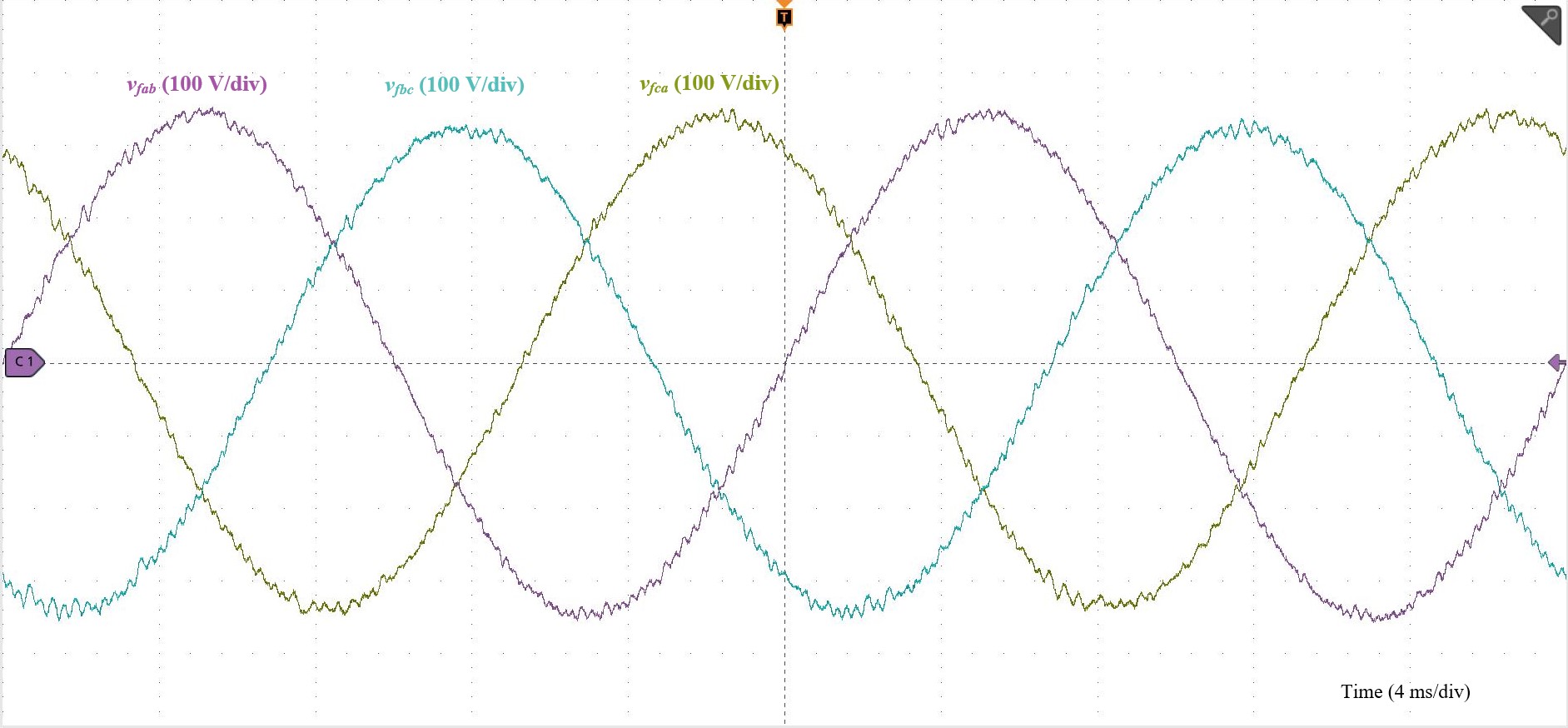}  
		\caption{Safe DQN-based controller. THD = 2.19\%.}
		\label{rl_control}
	\end{subfigure}
	\caption{Performance of different controllers.}
	\label{control_results}
 \vspace{-8mm}
\end{figure}

The safety-enhanced self-learning framework for optimal converter control is transferred from the edge device to a practical converter to experimentally validate the obtained optimal control policy with the deployment framework. The results shown in Fig. \ref{control_results} verify that the safety-enhanced self-learning optimal controller achieves desirable control performance with a THD of 2.19\%, comparable to the THD of 1.97\% for FCS-MPC with precise system parameters and established models. 

Moreover, sensitivity analysis regarding the parameter variations is performed. The variations in the model parameters would deteriorate the performance of FCS-MPC. In contrast, the proposed safe RL learning-based controller is model-free, and the proposed safe policy framework only functions when unsafe actions occur. Once converging within the safe learning region, as presented in Fig. \ref{current_tj}, the RL agent can find the optimal control strategy safely. A variation of inductance ($\triangle L_f$) and capacitance ($\triangle C_f$) within $\pm$ 30\% the nominal value is implemented to investigate the robustness of the safety-enhanced self-learning optimal control framework. 

If the safety framework underestimates the inductance value ($\triangle L_f > 0$), the safe exploration region is more tightly guaranteed as it reduces the current filter peaks and the reference current. On the other hand, if the safety framework overestimates the inductance value ($\triangle L_f < 0$), the filter current peaks may exceed the maximum value while still within the limits by properly leaving a margin between the predefined maximum current and physical limited current. As the training advances, the RL agent steadily converges within the designated safe region, where the current trajectories also become centered on the safe region.
In either case, the agent still finds the optimal control policy. Conversely, the capacitor is indirectly controlled by the inverter voltage due to the cross-coupling effect between the inductor and capacitor. Thus, the capacitance uncertainty barely influences the safe exploration of the RL agent. In summary, despite the parameter uncertainty, the proposed RL-based controller can still find the optimal switching policy for the converter.
\vspace{-2mm}
\section{Conclusions}
This paper proposes a safety-enhanced self-learning approach for optimal power converter control by introducing computational light single-step predictive control to the learning-based control framework. The proposed safe policy can guarantee the system's safety and reduce unnecessary exploration regions, thus also improving training efficiency. A deployment framework for transferring the online safe RL-based controller from simulation on edge devices to practical implementation in an experimental setup is also demonstrated. The experimental results confirm that the RL-based controller achieves a satisfactory control performance for VSC comparable to the conventional FCS-MPC. Future work will include self-learning control in real-time directly on the edge devices of a practical experimental setup. 
\vspace{-2mm}
\ifCLASSOPTIONcaptionsoff
  \newpage
\fi





\bibliographystyle{IEEEtran}
\bibliography{IEEEabrv,Bibliography}
\end{document}